\documentclass{jpsj2}
\title{Spin and charge orders and their hole-doping dependence in single layered cobaltate La$_{2-x}$Ca$_{x}$CoO$_{4}$ (0.3…$x$…0.8)}

\author{Kazumasa \textsc{Horigane}$^{1}$, Haruhiro \textsc{Hiraka}$^{2}$, Toru \textsc{Uchida}$^{1}$, Kazuyoshi \textsc{Yamada}$^{2}$ and Jun \textsc{Akimitsu}$^{1}$}

\inst{$^{1}$Department. of Physics and Mathematics, Aoyama-Gakuin University, Tokyo 229-8558 \\
$^{2}$Institute for Materials Research, Tohoku University, Katahira 2-1-1, Sendai, Miyagi 980-8577,Japan \\}

\abst{Neutron scattering experiments were performed on single crystals of layered cobalt-oxides La$_{2-x}$Ca$_{x}$CoO$_{4}$ (LCCO) to characterize the charge and spin orders in a wide hole-doping range of 0.3…$x$…0.8. For a commensurate value of $x$=0.5 in (H,0,L) plane, two types of superlattice 
reflections concomitantly appear at low temperature; one corresponds to a checkerboard charge ordered pattern of Co$^{2+}$/Co$^{3+}$ ions and the other is magnetic in origin. Further, the latter magnetic-superlattice peaks show two types of symmetry in the reflections, suggesting antiferromagnetic-stacking (AF-S) and ferromagnetic-stacking (F-S) patterns of spins along the $c$ direction. From the hole-doping dependence, the in-plane correlation lengths of both charge and spin orders are found to give a maximum at $x$=0.5. These features are the same with those of $x$=0.5 in La$_{1-x}$Sr$_{1+x}$MnO$_{4}$ (LSMO), a typical checkerboard and spin ordered compound. However, in (H,H,L) plane, we found a magnetic scattering peak at Q=(1/4,1/4,1/2) position below $T_{N}$. This magnetic peak can not be understood by considering the Co$^{2+}$ spin configuration, suggesting that this peak is originated from Co$^{3+}$ spin order. 
By analyzing these superlattice reflections, we found that they are originated from high-spin state of Co$^{3+}$ spin order. 
}

\kword{Cobalt oxide, Spin and charge order, Checkerboard charge order, Spin state of Co$^{3+}$, Neutron diffraction}

\begin{document}
\maketitle

\section{Introduction} %% No sections necessary for express letters, letters and short notes
Layered transition-metal oxides have attracted much attention due to their wide variety of magnetic, electrical and structural properties$^{1)}$. In some of the doped transition metal compounds, a real space ordering exists due to the doped charge carriers in a certain carrier concentration, resulting in a charge ordering (CO) and sometimes an orbital ordering (OO). The case of half-doped manganite La$_{0.5}$Sr$_{1.5}$MnO$_{4}$ (LSMO) has interesting aspects for several reasons. Sternlieb $et$ $al.$$^{2)}$ reported the superlattice reflections corresponding to an alternating Mn$^{3+}$/Mn$^{4+}$ checkerboard charge ordered pattern below $T_{co}$=210K. Below N\'{e}el temperature, $T_{N}$=110K, magnetic ordering was found to have a unit cell with dimensions $\sqrt{\mathstrut 2}a_{tet}$$\times$$\sqrt{\mathstrut 2}a_{tet}$$\times$$c$ relative to the chemical unit cell. The checkerboard order is thought to arise from orbital ordering on the Mn sites, and subsequently a zig-zag type orbital order, concomitant with CO, was directly observed in a single-layered manganite$^{3)}$ by resonant x-ray scattering. This CO-OO transition was found in many half-doped manganates. Half doped manganites with the small bandwidth and disorder which corresponds to the small and similar size of A site ions such as Pr$_{0.5}$Ca$_{1.5}$MnO$_{4}$ exhibit a long-range CO-OO$^{4)}$. This indicates that A-site ions are important to mediate the long-range CO-OO because of controlling the band-width and disorder by A-site ions. 

Recently, Zaliznyak $et$ $al.$$^{5)}$ investigated a checkerboard charge order of Co$^{2+}$/Co$^{3+}$ and Co$^{2+}$ magnetic order in the half-doped cobaltate La$_{1.5}$Sr$_{0.5}$CoO$_{4}$ (LSCO) by elastic neutron scattering. From the analysis of the structural and magnetic scattering, they concluded that a checkerboard arrangement of high-spin state Co$^{2+}$ and low-spin state Co$^{3+}$ ions was realized in the CoO$_{2}$ plane. Although these spin and charge configurations resemble those of LSMO, some physical characteristics are unclear. First, compared to the CO in isostructural LSMO, it has a much shorter correlation length ($\xi$$_{co}$(Co) = 26(2)\AA) than the LSMO one ($\xi$$_{co}$(Mn) $>$ 300\AA). Second, there is no information on charge and spin orders in a wide hole doping range. Finally, spin state of Co$^{3+}$ ion is not clarified. Moritomo $et$ $al$.$^{6)}$  reported magnetic susceptibility measurement and they concluded that high-spin state Co$^{2+}$ and Co$^{3+}$ were 
realized in La$_{1.5}$Sr$_{0.5}$CoO$_{4}$. Because this result is inconsistent with neutron experiment, Co$^{3+}$ spin state is yet controversial.  In this sense, a general understanding of charge and spin orders is needed in the layered cobalt oxide system. From this point of view, we extensively examined the Ca concentration (hole-doping) dependence of charge and spin orders in La$_{2-x}$Ca$_{x}$CoO$_{4}$.

More recently, we have succeeded in making the large single crystal in a wide hole carrier range (0.3…$x$…0.8) with layered cobalt system La$_{2-x}$Ca$_{x}$CoO$_{4}$ (LCCO) and investigated the change of the magnetic properties including their anisotropy$^{7)}$. We have observed significant reduction of Weiss temperature $\Theta$, accompanying reduction of effective moment $\mu_{eff}$ from 4.0$\mu_{B}$ to 3.0$\mu_{B}$ beyond $x$`0.7. Cobalt oxide system such as LaCoO$_{3}$$^{8)-13}$ is known to be a candidate material of Co$^{3+}$ intermediate spin (IS) state and these changes can be ascribed to a transition of the spin state of the Co$^{3+}$ from the high-spin state(HS) ($x$…0.5) to the intermediate-spin state ($x$†0.7). 

In this paper, we report on the elastic neutron scattering measurements of layered perovskite cobalt oxides La$_{2-x}$Ca$_{x}$CoO$_{4}$ (0.3…$x$…0.8). The Ca-doping brought about unexpected effects and new magnetic phases were formed. Magnetic scattering pattern suggested that it was isostructural to LSMO, however spin and charge ordered peaks coexist in a wide hole carrier range for LCCO, whereas the coexisting phase appears only in the vicinity of $x$=0.5 in LSMO. To verify these physical properties, we examined the spin/charge structure, A site substitution effects and its spin structure changes.

\section{Experimental}

Polycrystalline samples were prepared by solid-state reaction. First, stoichiometric mixtures of La$_{2}$O$_{3}$, CaCO$_{3}$, and CoO powders were ground and calcined at 1,200Ž for 24h in N$_{2}$ atmosphere. After repeating this grinding and sintering process three times, the powder sample was pressed into a rod with a size of 6 mm$\phi$~100mm and sintered at 1,100Ž for 24h. Single crystals of La$_{2-x}$Ca$_{x}$CoO$_{4}$ were grown by floating-zone method at a feeding speed of 3 mm/h in air. The ceramic compounds were confirmed to be a single phase using powdered portion by x-ray powder diffraction. The cylindrical crystals studied here for neutron scattering were about 5 mm in diameter and 30 mm in length. Typical FWHM mosaic was about 0.1‹. The lattice constants of $x$ = 0.5, for example, can be indexed in tetragonal notation of $I4/mmm$, however the crystal structure of $x$=0.5 is found to be orthorhombic (space group $A2mm$) at room temperature from recent convergent beam electron diffraction measurements. Thus, we use the orthorhombic unit cell (e.g., $a$`$b$`5.418ð and $c$`12.469ð for $x$=0.5).

Neutron scattering experiments were carried out on triple-axis spectrometers TOPAN and AKANE at JRR-3 in Japan Atomic Energy Agency, Tokai. Pyrolytic-graphite (PG) (0,0,2) reflection was used as a monochromator and an analyzer for TOPAN, and higher-order contaminations were removed by inserting two PG filters in the beam path. On the other hand, a combination of Ge (3,1,1) monochromator and PG (0,0,2) analyzer was employed for AKANE and no filters were attached owing to the forbidden Ge (6,2,2) reflection. The incident neutron energy of TOPAN [AKANE] was fixed to be $E_{i}$=30.5meV [19.6 meV], typically with a sequence of collimations of blank(50')-30'-S-60'-blank(180') [guide(20')-open-S-60'-blank(180')]. The sample was mounted in a $^4$He-cycle-type refrigerator, allowing wave-vector transfers in the (H,0,L), (H,H,L) scattering plane.

\section{Magnetic and charge order ($x$=0.5)}

In the previous LSCO neutron study$^{5)}$, magnetic peaks were observed at reciprocal lattice positions around (H,0,L) with H=half-integer and L=odd, while structural/charge ordered peaks were observed with H=odd and L=integer. In order to explore the charge and magnetic orders in the half-doped LCCO with $x$=0.5, we searched such the super-lattice structures in the (H,0,L) plane by using three trajectories from "scan A" to "scan C" as seen in the inset of Fig.1.

Before considering the actual data, it is essential to understand the distribution of Bragg points in the reciprocal-lattice plane because of the twin structure in the orthorhombic phase, the $a$* and $b$* axes are superposed. In the inset of Fig.1 we show the nuclear, structural and magnetic Bragg reflections in an $a$*($b$*)-$c$* reciprocal-lattice plane of the orthorhombic twin structure. The circle, triangle and square denoted the primary, structural and magnetic reflections, respectively, while filled and open figure means $A2mm$ and $Bm2m$ twin structure, respectively. 
 
Figure 1 [(H,0,1) or (0,K,1)] shows the "scan A" profiles below and above $T_{N}$ (=50 K). Since the half-integer peaks disappear above $T_{N}$, they should be magnetic scattering. On the other hand, the origin of the odd peaks must be nuclear (or charge) because the cross section is independent of temperature up to 71 K. Thus, a checkerboard type of charge order on Co$^{2+}$/Co$^{3+}$ is expected to realize in LCCO system. Figure 2 [(1,0,L) or (0,1,L)]  shows a Q spectrum of the "scan B" at 10 K. If our sample has only the single domain $A2mm$, we should observe the structural scattering for L=even. However, an integer-peak structure is clearly seen, indicating that our crystal has the twin structure. It is consistent with the checkerboard-type charge order observed in LSCO$^{5)}$ and LSMO$^{2)}$. Indeed, this order persists up to room temperature (not shown). However, this profile of LCCO is clearly different from that of LSCO system. In order to contrast to the LCCO and LSCO system,
 we fit the structural scattering spectra, a Lorenzian cross section was convoluted with the spectrometer resolution function using the program ResLib$^{14)}$. The correlation length of checkerboard charge order is $\xi$$_{ab}$=115(12)\AA in La$_{1.5}$Ca$_{0.5}$CoO$_{4}$ (LCCO), while $\xi$$_{ab}$=23(2) \AA in La$_{1.5}$Sr$_{0.5}$CoO$_{4}$$^{5)}$ (LSCO). The correlation length of LCCO is five times longer than that of LSCO. This result suggests that bandwidth and disorder can also be controlled by A site ion radius in layered cobaltates. On the other hand, L-dependence of magnetic scattering in LCCO looks somewhat complex. Figure 3 shows a magnetic scattering of La$_{1.5}$Ca$_{0.5}$CoO$_{4}$ in (0.5,0,L) [or(0,0.5,L)] (Scan C). Three sources of scattering can be distinguished; well ordered scattering at L=half integer, additional peaks centered at L=integer and broad component that corresponds to the magnetic diffuse scattering along L. From the temperature dependence of magnetic 
 peaks with L=integer and half-integer (Inset: Fig.3), it was found that the magnetic peaks with L=half-integer was sharply changed at $T_{N}$, while that of L=integer was gradually changed. Thus, it is clear that two magnetic peaks are obviously different in origin. Table I  shows the charge and magnetic correlation lengths in half-doped manganite$^{2),17)}$, nikelate$^{15)}$ and cobaltate$^{5),16)}$. The correlation lengths of LCCO obtained from magnetic peaks are $\xi_{ab}$=195(4)\AA, $\xi_{c}$=22(1)\AA (L=half-integer) and $\xi_{ab}$=98(8)\AA, $\xi_{c}$=11(1)\AA (L=integer), respectively. Comparing with the correlation lengths in half-doped layered perovskite, it was found that those of LCCO are the closest literature value for the magnetic and checkerboard charge order in LSMO.Moreover, the observed magnetic peaks of LCCO are in good agreement with previous neutron experiments for La$_{0.5}$Sr$_{1.5}$MnO$_{4}$$^{2),17)}$ and LaSr$_{2}$Mn$_{2}$O$_{7}$$^{18)}$. In LSMO, Larochelle $et$ $al.
$$^{17)}$ has also reported the two types of magnetic peaks and they concluded that the origin corresponds to two stacking patterns along the [001]. The type-I (type-I\hspace{-.1em}I) stacking pattern has magnetic peaks at half-integer (integer) L positions, corresponding to antiferromagnetic (ferromagnetic) next-NN planes. According to the two stacking model, a weakly L-dependent diffuse scattering may appear due to stacking fault of type-I and type-I\hspace{-.1em}I spin structure. And we can explain long-ranged magnetic correlation length at L=half-integer by considering the majority type-I and minority type-I\hspace{-.1em}I domains. This model can explain the magnetic scattering position, however this model can not explain the reason why two types of stackings are existed. Thus, we took into account the four magnetic domains by considering the two independent spin configuration and two types twin structure. Figure 4 shows the schematic spin and charge configuration model in La$_{1.5}$Ca$_{0.5}$CoO$_{4}$. Four magnetic domains are realized by considering the spin configuration and twin structure. As a stacking, we considered that there is a stacking vector along the [0,1/2,1/2] direction because $A2mm$ has two equivalent site, (0, 0, 0) and (0,1/2,1/2). As LCCO has the twin structure, the stacking vector should be along the two directions. Therefore, magnetic peaks with L=half-integer (integer) can be explained by B-I(A-I\hspace{-.1em}I) magnetic domain in (0.5,0,L). In brief, each magnetic reflection is contributed from about 25$\%$ magnetic domains. We analyzed the spin structure based on this model as discussed later.

Next, we searched new super-lattice reflections in the (H,H,L) plane. Figure 5-(a),(b) shows magnetic scattering scans along Q=(H,H,0.5)(ScanD) and (0.75,0.75,L)(ScanE), respectively. The profile of the magnetic scattering exhibits peaks at H = (2n+1)/4, l=m and m/2 positions, here, n and m are integers. In this case, sharp magnetic peaks at L=half-integer and broad magnetic peaks at L=integer were observed, suggesting that two types of magnetic peaks are strongly correlated with magnetic stacking along the $c$-axis. The presence of (0.25,0.25,1) and (0.25,0.25,0.5) peaks in these spectra indicates that the magnetic unit cell has dimensions $4a_{ortho}$$\times$$4a_{ortho}$$\times$$c$ (or 2$c$) relative to the orthorhombic unit cell. This magnetic unit cell can not be understood by only considering the Co$^{2+}$ spin configuration. Moreover, the in-plane correlation length for L=half-integer and integer magnetic order are 93(8)\AA, 74(4)\AA, respectively. These correlation lengths are 
much shorter than that of Co$^{2+}$ spin order. Thus, we concluded that the new magnetic peaks are originated from the magnetic order of Co$^{3+}$ spins.

\section{Hole-doping dependence of magnetic and charge orders (0.3…$x$…0.8)}

Next, we move onto the hole-doping dependence of the charge and magnetic orders in a wide doping range of 0.3…$x$…0.8. Figure 6-(a) shows the charge order spectrum of (1,0,L) [or (0,1,L)] for $x$=0.4, 0.5 and 0.8. Charge order peaks are insensitive to $x$, indicating a robust checkerboard-type charge order. On the other hand, this hole-doping process drastically changes the magnetic diffraction pattern below and above $x$=0.5. Magnetic scattering at (0.5,0,L)[or(0,0.5,L)] for $x$=0.4 and 0.8 are shown in Fig.7-(a),(b). The type-I reflections (L=half-integer) appear below $x$=0.5 together with the broad diffuse scattering, whereas it switches to the type-I\hspace{-.1em}I reflections (L=integer) with a flat background in the high doping range of $x$$>$0.5. This result is consistent with the Larochelle's result of perovskite manganites. From the results, we can estimate the charge and magnetic correlation lengths in LCCO system. Table I\hspace{-.1em}I shows the Ca dependence of charge and magnetic correlation lengths. In the low doping range of $x$$<$0.5, the charge and magnetic correlation lengths were rapidly decreased with decreasing $x$. On the other hand, charge correlation lengths of high doping range were gradually decreased with increasing $x$. Particularly, magnetic correlation lengths tend to increase with increasing $x$. This tendency suggests that charge and spin orders are strongly correlated with each other. We discuss the strange Ca dependence of magnetic correlation lengths in a next section.

Comparing the results for LCCO with those of the half-doped perovskites, we concluded that the spin and charge configurations in the present system were almost similar with those of LSMO. However, some features of LCCO are different from that of LSMO system. First is the magnetic structure. Neutron and x-ray diffraction measurements revealed that CE-type charge/spin ordering is realized in LSMO system. In (H, H, L)$_{ortho}$ zone, the magnetic reflections originated from Mn$^{4+}$ site in the CE-type spin ordering can be observed at the ((2n+1)/2,(2n+1)/2),m/2) positions, here, n and m are integers, respectively. On the other hand, magnetic scattering due to the Co$^{3+}$ exhibits peaks at H = (2n+1)/4, L=m and m/2 position. These observed peaks are obviously different from that of the CE-type spin structure and it is necessary to determine the spin structure for LCCO system. Second, in the entire $x$ range searched for LCCO, charge and spin ordered peaks coexist, whereas the 
coexisting phase appears only in the vicinity of LSMO. Third is the Ca concentration dependence of charge order. Long-range checkerboard charge order is present in the wide Ca concentration range and no incommensurability was observed in all cases. On the other hand, charge ordering disappears at $x$=0.45 in LSMO. Moreover, at doping $x$$>$0.5, long-range structural order is present and this order is incommensurate with the lattice. To understand these LCCO characters, we discuss the hole carrier mobility, A site substitution effects and spin structure analysis

\section{discussion}

Commensurability of CO is present in LCCO system at all doping range, while this order is incommensurate in LSMO system. One of the possible interpretation for their difference is hole carrier mobility. The resistivity of LCCO system is insulating over the whole temperature range. By contrast, the resistivity of LSMO is considerably small ($\rho$`2$\times$10$^{-1}$ƒ¶$\cdot$cm at 300K). The difference of resistivity can be ascribed to the bandwidth, which of LSMO is relatively larger than that of LCCO. Thus, it is expected that charge and magnetic orderings are suppressed by Sr doping in LSMO, and lead to the incommensurability of charge ordered peaks. On the other hand, electrons are localized in LCCO and commensurate charge and magnetic orderings are observed in a wide $x$ range. The other possibility is the relationship between incommensurability $\epsilon$ and $e_{g}$ electron density $n_{e}$. In LSMO system, Larochelle reported that incommensurability $\epsilon$ shows the $\epsilon
$`2$n_{e}$ law at doping range $x$$>$0.5. Applying this model to LCCO system, commensurate CO can be understood by assuming high-spin state of Co$^{2+}$ and Co$^{3+}$.

We considered that the correlation length difference between LCCO and LSCO is due to the bandwidth and disorder of A site ions. In the layered manganite case$^{4)}$, Mathieu $et$ $al.$ estimated the bandwidth and disorder by the average A-site ionic radius ($r_A=\sum_{i}x_ir_i$) and the ionic radius variance ($\sigma^2=\sum_{i}x_ir_i^2-r_A^2$), where $x_{i}$ and $r_{i}$ are the fractional occupancies ($\sum_{i}x_i=1$) and electronic radii of the different $i$ cations on the A site, respectively. In LCCO system, $r$$_{A}$ and $\sigma$$^2$ are very small value ($r$$_{A}$`1.15\AA, $\sigma$$^2$`3$\times$10$^{-4}$\AA$^2$) while, those of LSCO are larger ($r$$_{A}$`1.19\AA, $\sigma$$^2$`2$\times$10$^{-3}$\AA$^2$). Therefore, Ca system has a smaller bandwidth and disorder than that of Sr system and the long-ranged CO are expected in LCCO. Thus, we concluded that the difference of charge correlation length between Ca and Sr can be explained by the bandwidth and disorder of the A site ions. 

Figure 7-(a),(b) indicate that magnetic scattering due to the type-I stacking is dominant for a low doping region ($x$$<$0.5), while that of type-I\hspace{-.1em}I stacking is dominant for a high doping ($x$$>$0.5). And broad L-dependent peaks are observed in a low doping region and maximized at $x$=0.5, however these peaks are suppressed in a high doping region. From these results, we concluded that the type-I and I\hspace{-.1em}I stacking domains are changed by Ca concentration region. The boundary of type-I and I\hspace{-.1em}I domains is expected to be at $x$=0.5 because two-types of magnetic scatterings are dramatically changed in a narrow region at $x$=0.5. From this finding, we explain the Ca dependence of broad L-dependent peaks and magnetic correlation lengths. In the low doping region, broad L-dependent peaks are observed because two-magnetic domains are formed by Ca concentration. In particular, the broad peak at $x$=0.5 is stronger than the other Ca concentration due to the
 commensurate cobalt valence Co$^{2.5+}$. In a high doping region, broad L-dependent peaks are suppressed because magnetic ordering is destroyed by hole doping, accompanying a reduction of type-I. Moreover, type-I\hspace{-.1em}I stacking domain is increased with Ca concentration in this model, suggesting that magnetic correlation length of type-I\hspace{-.1em}I stacking is increased with Ca doping. In order to support our consideration, we studied the magnetic scattering in oxygen-excessive perovskite La$_{1.85}$Ca$_{0.15}$CoO$_{4.17}$. If our consideration is correct, it is that expected type-I stacking domain peaks are only observed in La$_{1.85}$Ca$_{0.15}$CoO$_{4.17}$, of which cobalt valence is about Co$^{2.5+}$. Figure 8-(a) shows magnetic scattering in La$_{1.85}$Ca$_{0.15}$CoO$_{4
.17}$. We observed the strong intensity peaks at L=half-integer and no magnetic diffuse peaks in La$_{1.85}$Ca$_{0.15}$CoO$_{4.17}$. This result indicates that type-I stacking domain is dominant in a lower doping, suggesting that the two domains are changed by Ca concentration at around $x$=0.5. However, this finding does not mention why magnetic structure dramatically changes by Ca concentration. To understand this, we should study other cobaltates which are substituted at A-site with various rare-earth and alkaline earth metals.   

Finally, we analyzed the magnetic scattering due to Co$^{2+}$ spin order in the (H,0,L) plane to determine the magnetic structure. In our analysis, we assumed that the magnitude of all Co-moments at crystallographically equivalent site was equal. We used the isotropic Co$^{2+}$ magnetic form factor which was reported in ref.19). For the spin direction, we used the following equation
\begin{eqnarray}
sin^{2}\Phi_s+cos^{2}\Phi_s\cdot sin^2\theta,
\end{eqnarray} 
where $\theta$ is the angle between the scattering vector and the CoO$_{2}$ plane. $\Phi_s$ is the angle between the spins and the (H,0,0) [(0,K,0)] axis in domain A (B). In our magnetic susceptibility measurements, we found that the magnetic easy axis is within the $ab$-plane. Thus, spins lie primarily in the basal plane. The integrated intensities of the magnetic scattering were obtained by taking accounts of both magnetic ordered and diffuse scatterings. The Lorentz factor-corrections were made.

Table I\hspace{-.1em}I\hspace{-.1em}I shows the comparison between the observed and calculated magnetic intensities at 9K in (H,0,L) plane. Comparing between the integrated intensities ($I$$_{obs}$) and the model calculation ($I$$_{cal}$), we obtained about four times larger calculated intensities than those of observed ones. This result indicates that we must consider the magnetic volume fraction due to the four magnetic domains A-I, A-I\hspace{-.1em}I, B-I and B-I\hspace{-.1em}I(see Fig.4). As mentioned above, we assumed that the observed magnetic reflections represent following equation.
\begin{eqnarray}
I_{obs}=r_{B-I}\cdot I_{cal}(type-I)+r_{A-I\hspace{-.1em}I}\cdot I_{cal}(type-I\hspace{-.1em}I).
\end{eqnarray} 

Here, $r$ is the volume fraction of magnetic domains and we defines $r$(A-I\hspace{-.1em}I) as 0.5 - $r$(B-I). We can qualitatively explain the results by assuming $r$(B-I) =0.315 in four domain model, indicating that the analyzed magnetic structure is essentially correct. This magnetic volume fraction $r$ is consistent with the assumption that type-I magnetic domain is dominant. Since the obtained magnetic moment of the Co$^{2+}$ is 2.86(19)$\mu_{B}$, which is almost as same as the value in a high-spin state of Co$^{2+}$, we concluded that Co$^{2+}$ is in the HS state. 

In order to confirm reasonability of the analysis, we analyzed the spin structure of La$_{1.85}$Ca$_{0.15}$CoO$_{4.17}$ which has no diffuse peaks and commensurate cobalt valence of Co$^{2.5+}$. The results are shown in Table I\hspace{-.1em}V. In the case of $r$(B-I)=0.50, we can qualitatively explain the result, which indicates that the obtained magnetic structure is essentially correct. By comparing the results between La$_{1.5}$Ca$_{0.5}$CoO$_{4}$ and La$_{1.85}$Ca$_{0.15}$CoO$_{4.17}$, we found that the magnitude $\mu$ of the Co$^{2+}$ moment and the angle $\Phi$$_{s}$ of $x$=0.5 are almost same as those of $x$=0.15, and this shows that our analysis is reasonable.  

Next, we discuss the features of Co$^{3+}$ magnetic scattering. To determine the spin state of Co$^{3+}$, we analyzed the magnetic scattering due to Co$^{3+}$ spin order in the (H,H,L) plane by assuming the magnetic structure as shown in Fig.9. We assumed that the alignment of spin moments is ªª«« type in CoO$_{2}$ plane. Table V shows the comparison between the observed and calculated magnetic intensities at 12K in (H,H,L) plane. By using the same magnetic volume fraction of $r$=0.315, we can explain the magnetic scattering in (H,H,L) plane. The magnitude ƒÊof the Co$^{3+}$ moments is 3.64(23)$\mu_{B}$, suggesting that Co$^{3+}$ is in the HS state. This result is consistent with our magnetic susceptibility measurements, and suggests that magnetic peaks at Q=(1/4,1/4,$\ell$) also observed in LSCO system. Indeed, we observed the magnetic peaks at Q=(1/4,1/4,$\ell$)($\ell$=integer) in La$_{1.5}$Sr$_{0.5}$CoO$_{4}$. By taking into account these new superlattice reflections, we will explain the inconsistency between neutron and magnetic susceptibility measurements in LSCO.

As is shown in Fig.9, we have found ª©«¨ªspiral order pattern along the [110]$_{ortho}$ direction. To stabilize this spiral structure, it is necessary to introduce the magnetic frustration. For Co$^{2+}$ spins, static spin ordering is robust because magnetic ordering was observed at all Ca concentrations. On the other hand, Co$^{3+}$ spins are weakly ordered because no magnetic ordering is observed for $x$=0.4, 0.6 (not shown). This physical picture can be easily understood by introducing spin frustration between Co$^{2+}$ and Co$^{3+}$ spins. One of the possible interpretation to realize this spiral structure is the competition of exchange interactions. However, it is not so straightforward to understand why the ª©«¨ª spiral ordering is realized below $T_{N}$. The clue to this problem might be obtained by the future study on the Ca concentration dependence in the LCCO system.

\section{conclusion}

We have studied various Ca concentration of magnetic and charge orders in LCCO by elastic neutron diffraction. By comparing the LSCO system, LCCO exhibits long-ranged magnetic and checkerboard charge orderings. This difference can be explained by the bandwidth and disorder of A-site ions. This result also indicates the importance of the bandwidth and disorder by A-site ion radius in layered cobalt oxides. At $x$=0.5, we found the structural and two-types of magnetic scatterings in (H,0,L) plane. In the magnetic structure analysis, we can qualitatively explain the neutron scattering results by four domain model.  The magnitude ƒÊ of the Co$^{2+}$ moment was 2.86(19)$\mu_{B}$ and we concluded that Co$^{2+}$ is in the HS state. For magnetic scattering in (H,H,L) plane, these magnetic scattering intensities are originated from Co$^{3+}$ spin order because this peak can not be understood by the Co$^{2+}$ spin configuration. The magnitude ƒÊof the Co$^{3+}$ moments is 3.64(23)$\mu_{B}$, 
suggesting that Co$^{3+}$ is also in the HS state. We think that this finding will explain the inconsistency between neutron and magnetic susceptibility measurements in LSCO.

For hole-doping dependence, we observed robust commensurate charge ordered peaks in a wide doping range. Although charge ordered peaks are insensitive to $x$, magnetic scattering pattern are drastically changed at $x$=0.5. The type-I reflections appear for $x$$<$0.5, whereas it switched to the type-I\hspace{-.1em}I reflections for $x$$>$0.5. Our results for doping dependence of magnetic and charge orderings give a refined understanding of the phase diagram of layered cobaltates. However, it is not clear why the magnetic spin structures are dramatically changed at $x$=0.5. Moreover, hole-doping dependence of magnetic scattering for Co$^{3+}$ is not so well understood. To clarify the clues, the investigation of the hole-doping effect is now in progress by our group. 

\section*{Acknowledgment}
The author acknowledges invaluable discussions with Profs. K. Asai, M. Suzuki and K. Abe. This work was partly supported by the Iwanami Fujukai Foundation and the 21st COE program, "High-Tech Research Center" Project for Private Universities: matching fund subsidy from MEXT (Ministry of Education, Culture, Sports, Science and Technology; 2002-2004), and a Grant-in-Aid for Scientific Research on Priority Area from the Ministry of Education, Culture, Sports, Science and Technology of Japan.

\begin{figure}[c]
\begin{center}
\includegraphics[width=14cm, keepaspectratio]{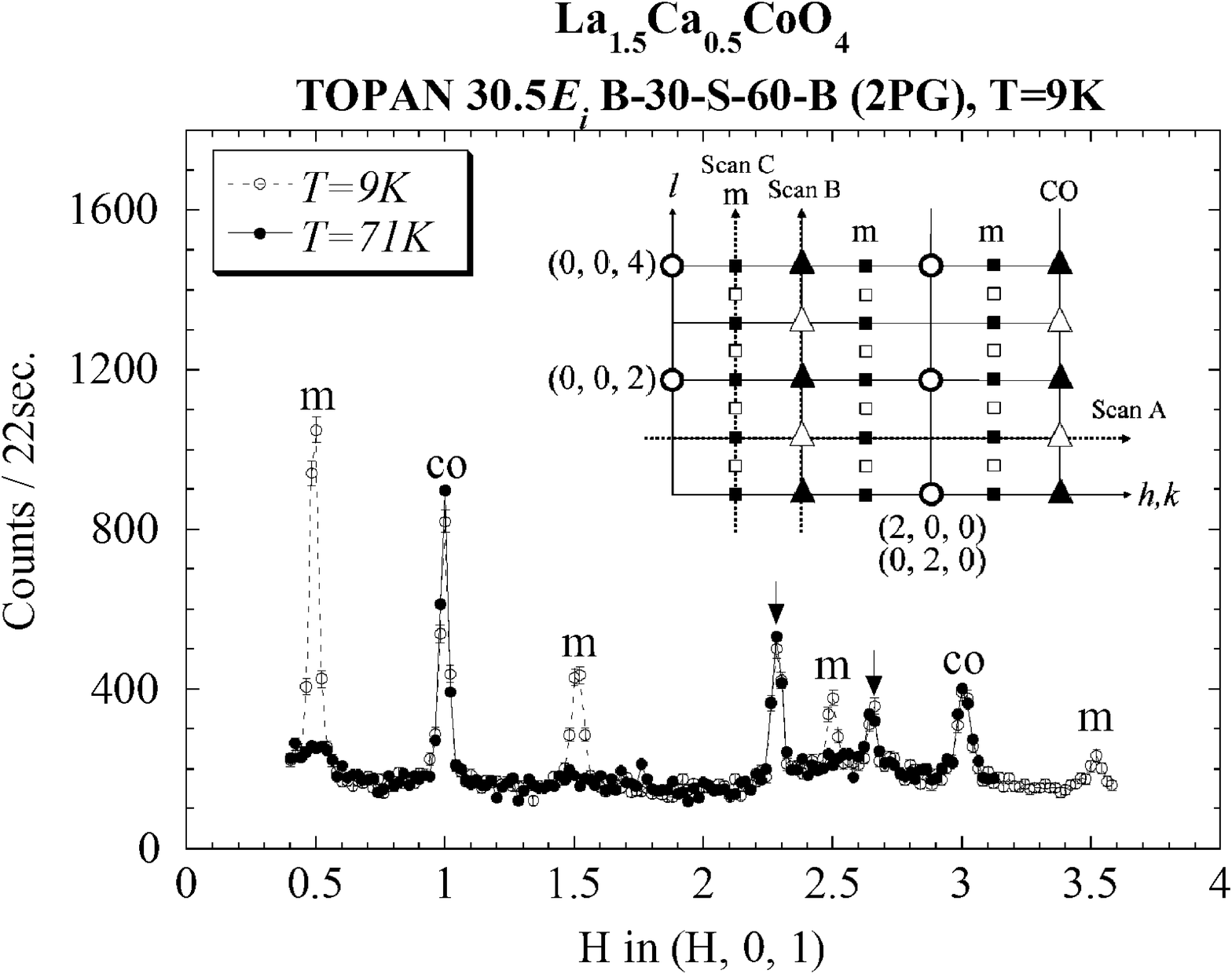}
\end{center}
\caption{Magnetic(denoted as $m$) and charge ordered(CO)/structural scatterings at (H,0,1) [ or (0,K,1)] in La$_{1.5}$Ca$_{0.5}$CoO$_{4}$ (Scan A). The peaks at H=half integer result from the magnetic scattering, while the peaks at H=odd are due to the structural modulation of checkerboard charge order. The arrows indicate the aluminum scattering. Inset: $a$*($b$*)-$c$* reciprocal-lattice plane for La$_{1.5}$Ca$_{0.5}$CoO$_{4}$. The circle, triangle and square denote the primary, structural and magnetic reflections, respectively, while filled and open figure means $A2mm$ and $Bm2m$ twin structure, respectively.}
\label{f1}
\end{figure}
\clearpage
\begin{figure}[c]
\begin{center}
\includegraphics[width=14cm]{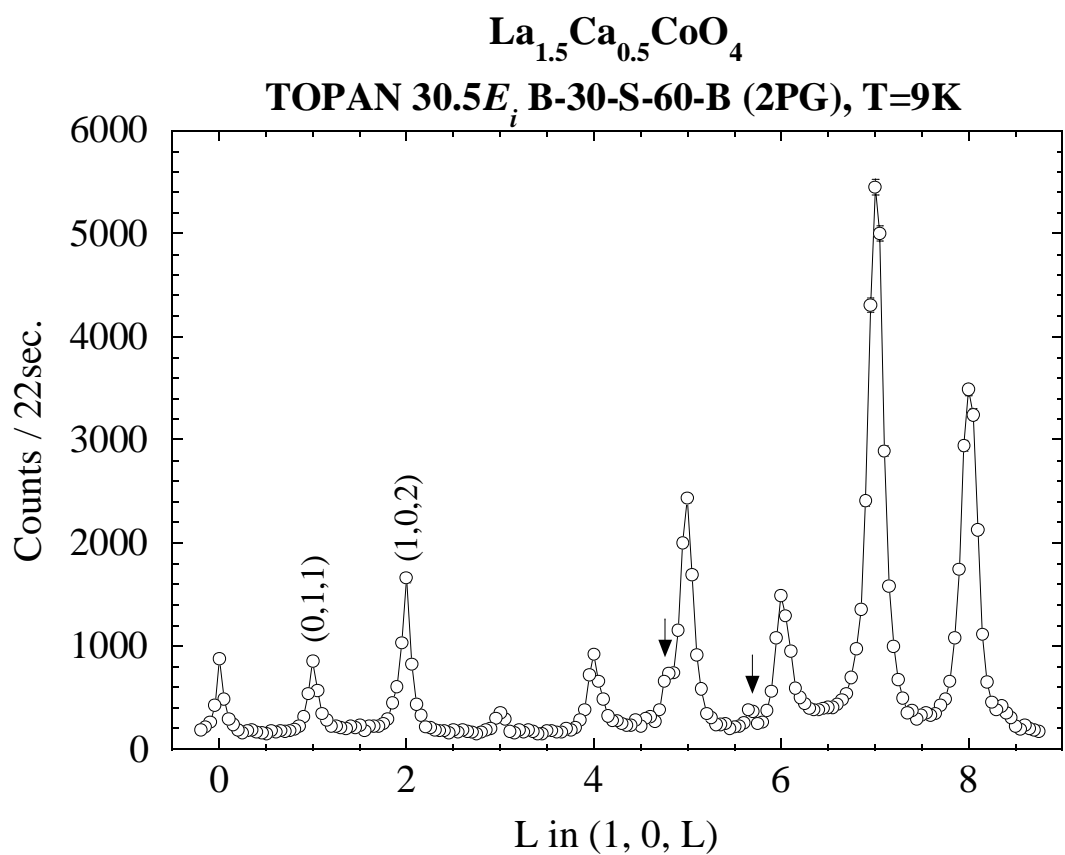}
\caption{Structural scattering due to the checkerboard charge order at (1,0,L)[or (0,1,L)] in La$_{1.5}$Ca$_{0.5}$CoO$_{4}$ (Scan B). The integer-peaks reflect the twin structure in our crystal. The arrows indicate aluminum scattering.}
\label{f1}
\end{center}
\end{figure}
\clearpage
\begin{figure}[c]
\begin{center}
\includegraphics[width=14cm]{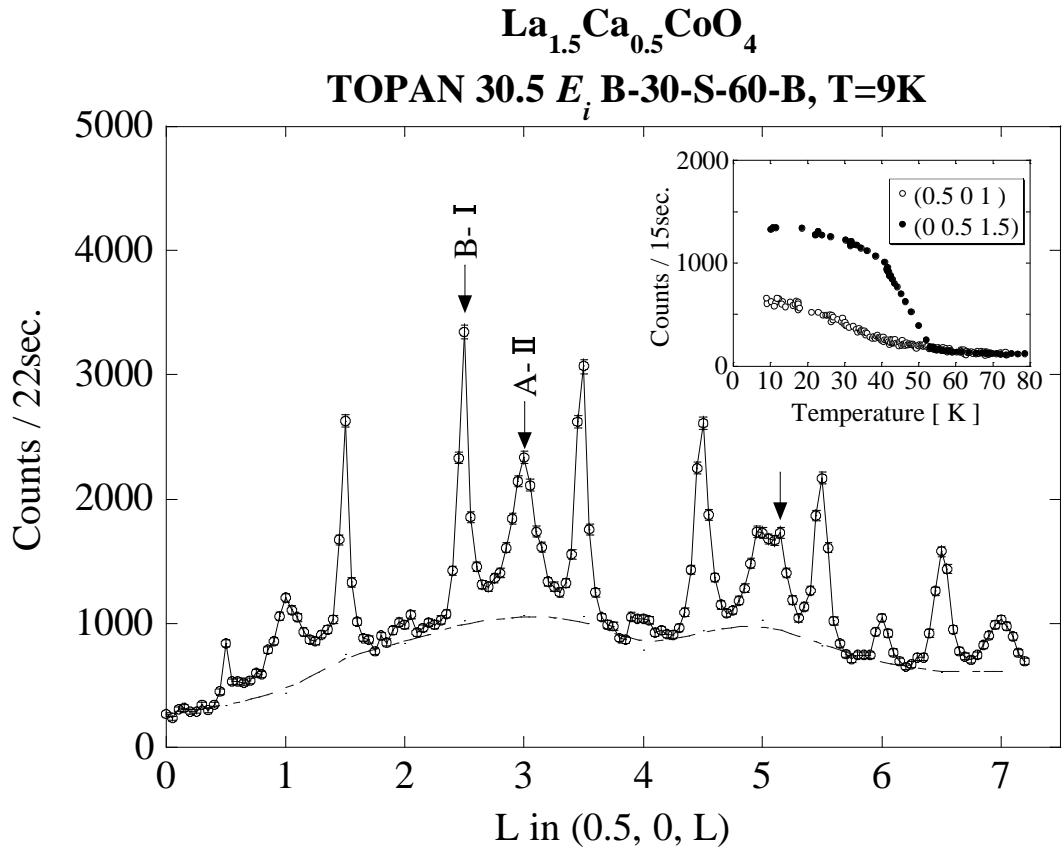}
\end{center}
\caption{Magnetic peaks at (0.5,0,L) [or(0,0.5,L)] in La$_{1.5}$Ca$_{0.5}$CoO$_{4}$(Scan C). Sharp magnetic peaks with L=half-integer (B-I) and broad magnetic peaks with L=integer (A-I\hspace{-.1em}I) are observed in (0.5,0,L). Broken line indicates a magnetic diffuse scattering (guide to the eye). Inset: temperature dependence of magnetic peaks with L= integer and half integer in La$_{1.5}$Ca$_{0.5}$CoO$_{4}$.}
\label{f1}
\end{figure}
\clearpage
\begin{figure}[c]
\begin{center}
\includegraphics[width=14cm]{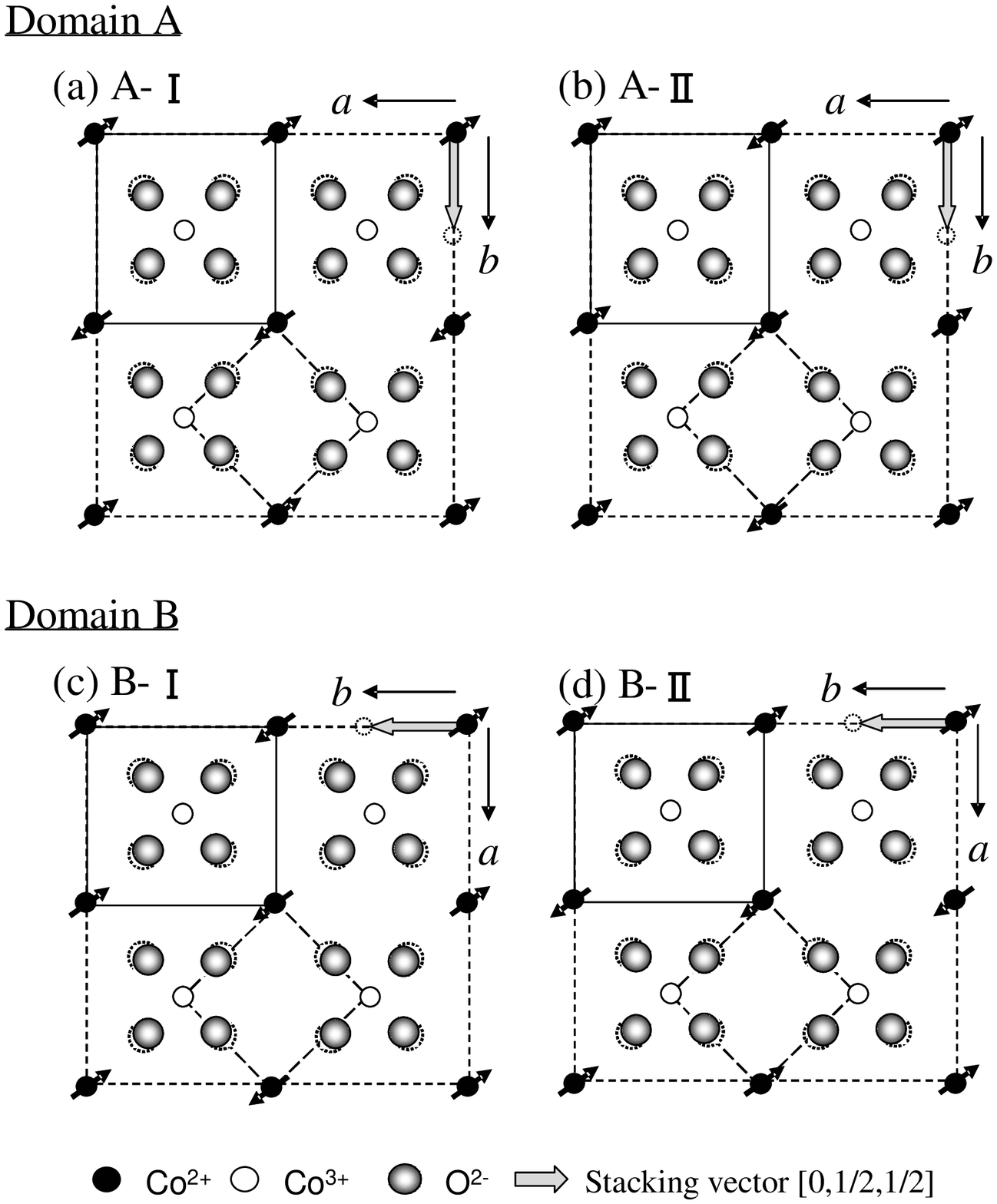}
\end{center}
\caption{Schematic spin and charge configuration model in La$_{1.5}$Ca$_{0.5}$CoO$_{4}$. Domains A, B represent $A2mm$ and $Bm2m$ twin structure, respectively. Four magnetic domains are realized by considering spin configuration and twin structure. The magnetic reflections can be interpreted by the type-I and type-I\hspace{-.1em}I stacking vectors as shown in the figure.}
\label{f1}
\end{figure}
\clearpage
\begin{figure}[c]
\begin{center}
\includegraphics[width=12cm]{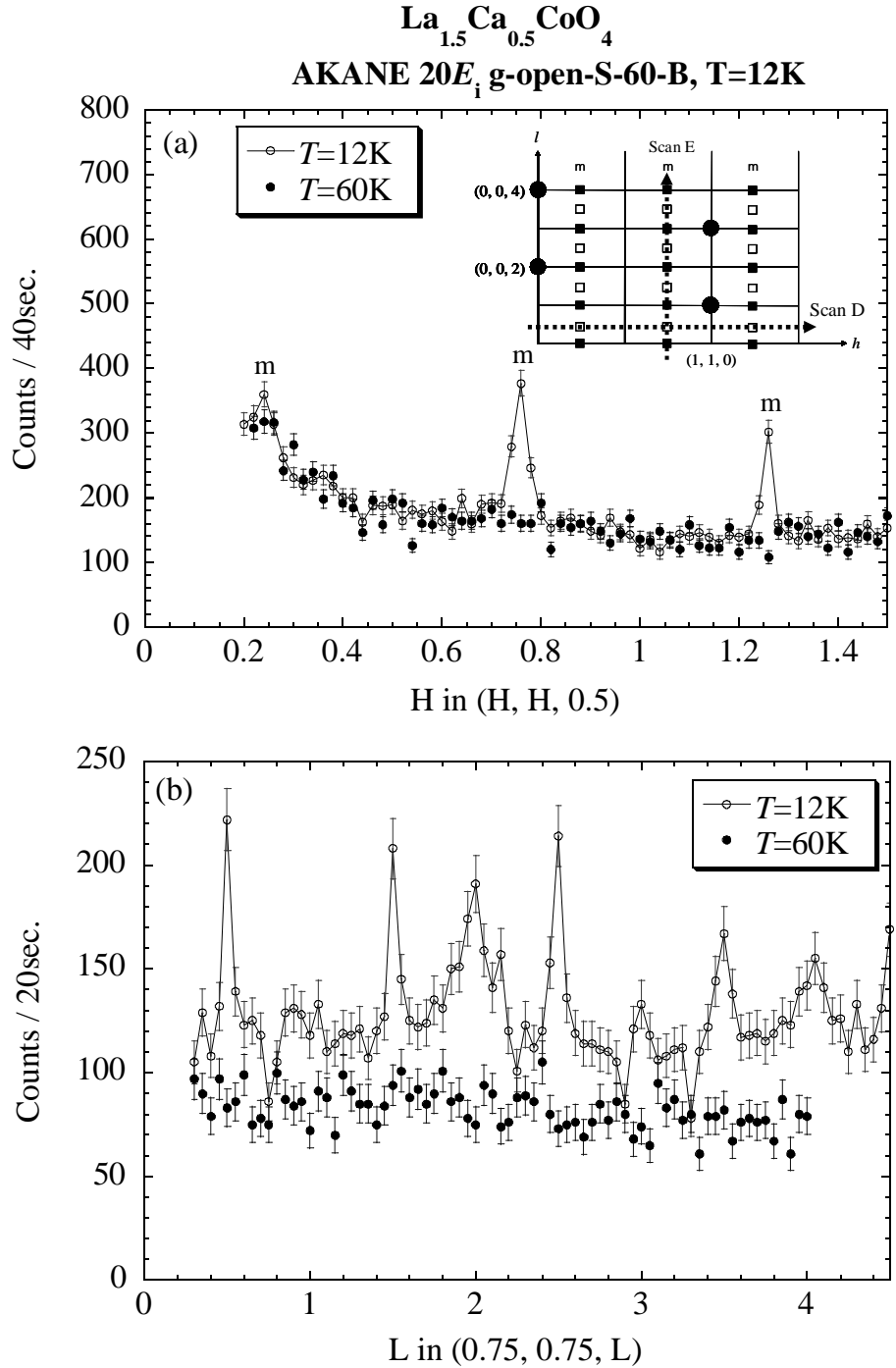}
\end{center}
\caption{(a)Magnetic scattering scans along (H,H,0.5)(Scan D) and (b) (0.75,0.75,L)(Scan E), respectively. The profile of the magnetic scattering exhibits peaks at H=(2n+1)/4, L=m and m/2 position, here, n and m are integers. Inset: (H, H, L) reciprocal plane for La$_{1.5}$Ca$_{0.5}$CoO$_{4}$. Filled and open squares mean L=integer and half-integer magnetic scatterings, respectively. Filled circles are nuclear reflections.}
\label{f1}
\end{figure}
\clearpage
\begin{figure}[c]
\begin{center}
\includegraphics[width=10cm]{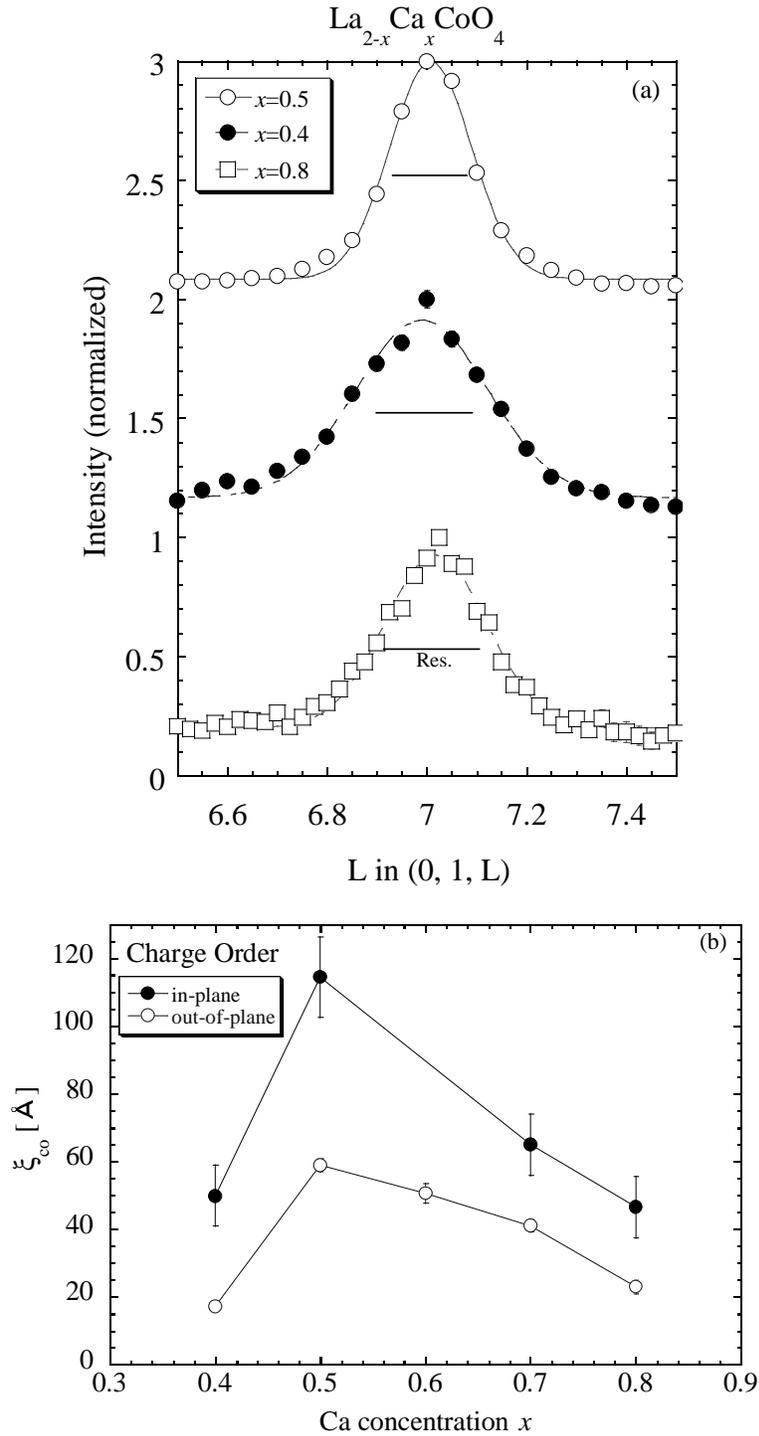}
\end{center}
\caption{(a) L scans at $T$=13K through the (1,0,L) structural scattering for $x$=0.5, 0.4, 0.8, normalized at the peak positions. (b) Ca dependence of the charge ordered correlation length in La$_{2-x}$Ca$_{x}$CoO$_{4}$. The correlation lengths were extracted from the scans by assuming a Lorentzian cross section convoluted with the instrumental resolution.}
\label{f1}
\end{figure}
\clearpage
\begin{figure}[c]
\begin{center}
\includegraphics[width=9cm]{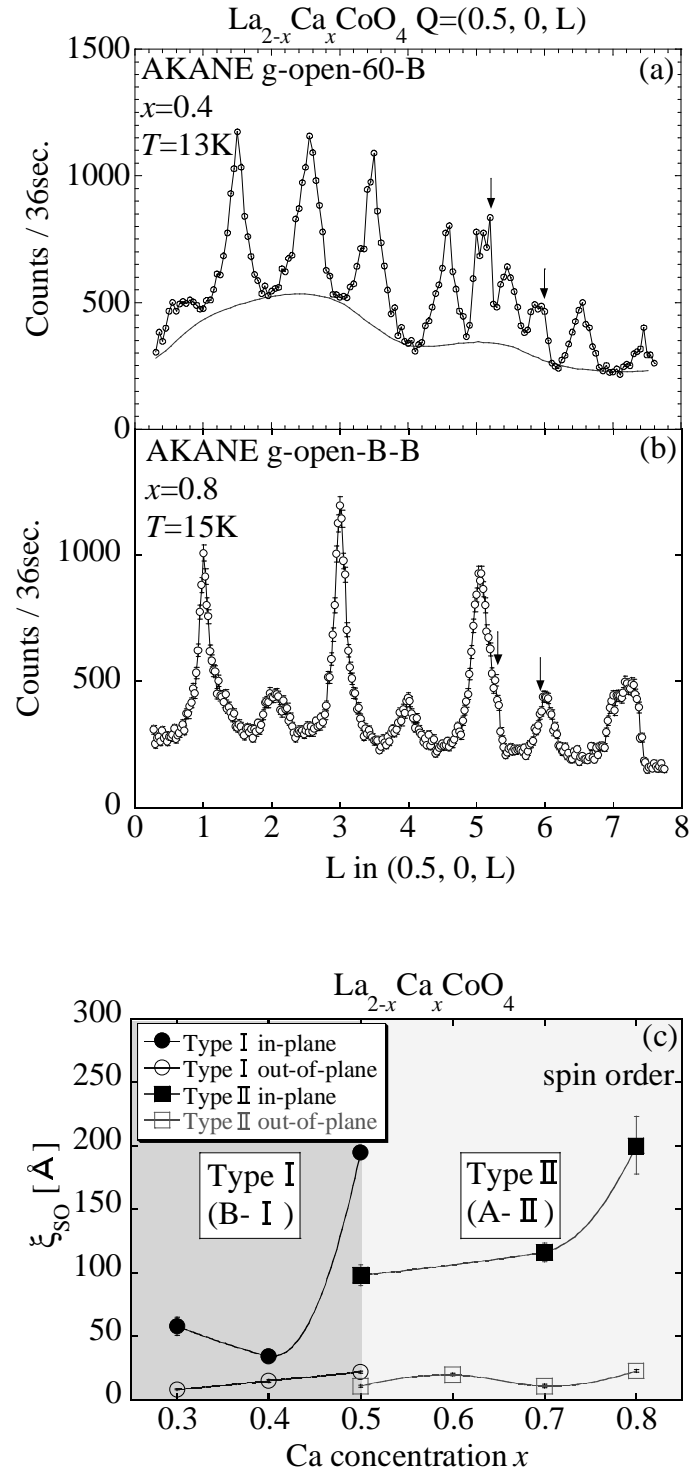}
\end{center}
\caption{(a),(b) Ca dependence of magnetic scattering at (0.5,0,L) [or (0,0.5,L)] for $x$=0.4 and $x$=0.8. In a lower doping region ($x$$<$0.5), type-I stacking domain (L=half-integer) was observed. While in a higher doping region ($x$$>$0.5), type-I\hspace{-.1em}I stacking domain (L=integer) was observed. Broken line indicates a magnetic diffuse scattering (guide to the eye). (c) Ca dependence of magnetic correlation length. The correlation lengths were extracted from the scans by assuming a Lorentzian cross section convoluted with the instrumental resolution.}
\label{f1}
\end{figure}
\clearpage
\begin{figure}[c]
\begin{center}
\includegraphics[width=12cm]{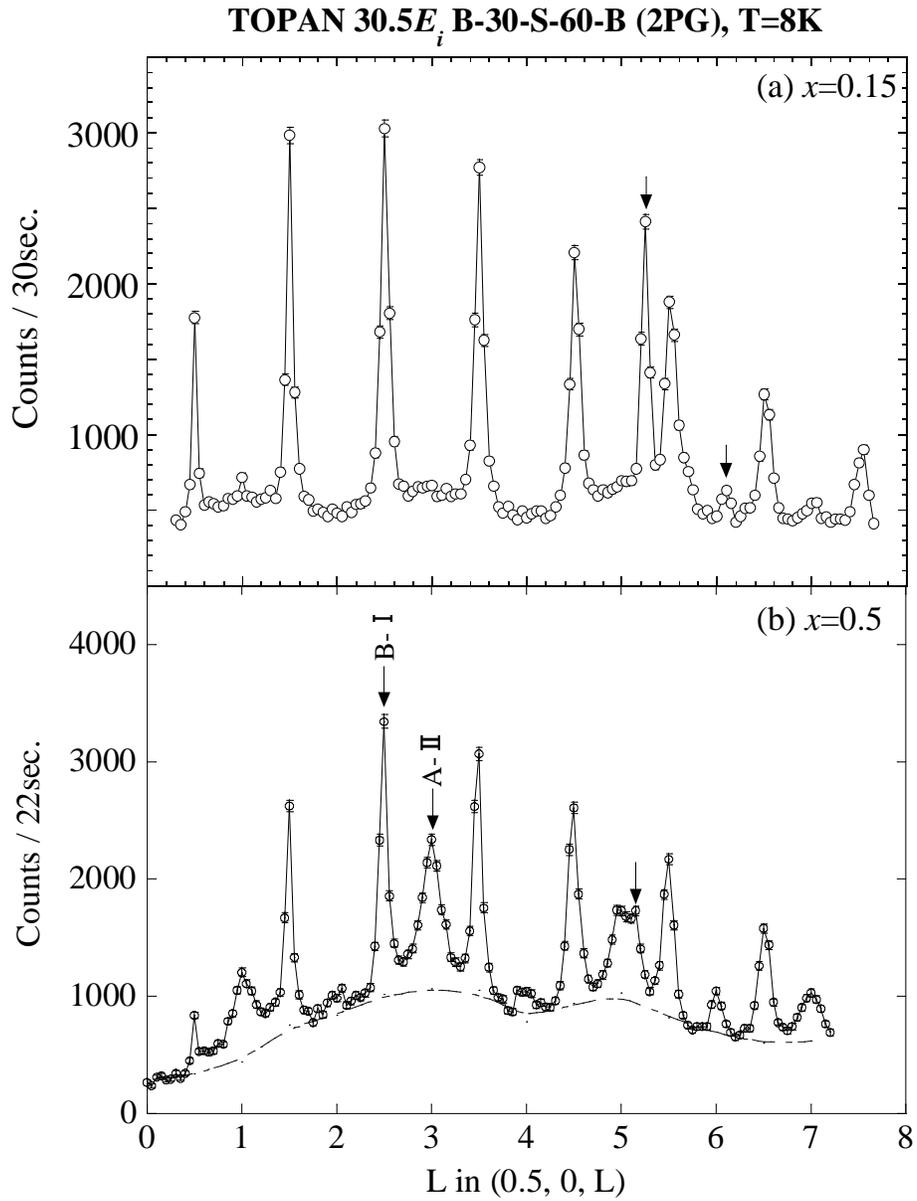}
\end{center}
\caption{(a) Magnetic scattering in La$_{1.85}$Ca$_{0.15}$CoO$_{4.17}$ . For $x$=0.15, strong type-I magnetic scattering and no diffuse scattering were observed. (b) same scan profiles for $x$=0.5.}
\label{f1}
\end{figure}
\clearpage
\begin{figure}[c]
\begin{center}
\includegraphics[width=12cm]{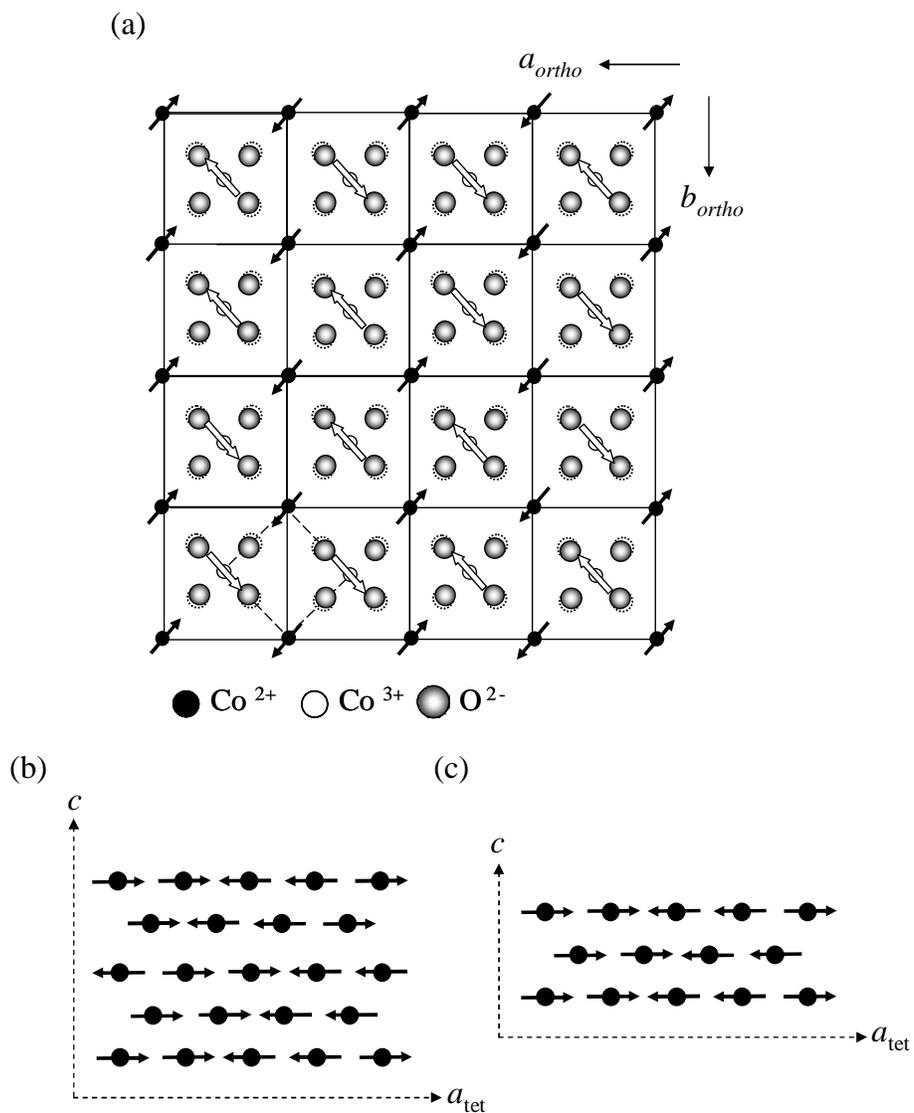}
\end{center}
\caption{(a) Co$^{2+}$ and Co$^{3+}$ spin arrangement in the CoO$_{2}$ plane. Possible Co$^{3+}$ spin arrangement model (b) antiferomagnetic and (c) ferromagnetic coupling pattern along the c-axis.}
\label{f1}
\end{figure}
\clearpage
\begin{table}[c]
\caption{Charge and magnetic correlation lengths in half-doped layered perovskites. In La$_{1.5}$Ca$_{0.5}$CoO$_{4}$(LCCO), $\xi$$_{co}$ and $\xi$$_{so}$ were determined by the (0,1,7), (0,0.5,2.5) and (0.5,0,1) peaks. While, those of La$_{1.5}$Sr$_{0.5}$CoO$_{4}$(LSCO)[5,16], La$_{1.5}$Sr$_{0.5}$NiO$_{4}$ (LSNO)[15] and La$_{0.5}$Sr$_{1.5}$MnO$_{4}$(LSMO)[2,17] were determined by all charge and magnetic peaks.}
\label{t1}
\begin{tabular}{cccccccc}
\\
\hline
 &\multicolumn{2}{c}{$ $}&\multicolumn{2}{c}{$L=half-integer$}&\multicolumn{2}{c}{$L=integer$}\\
 &$\xi_{co}^{ab}$ (\AA)&$\xi_{co}^{c}$ (\AA)&$\xi_{so}^{ab}$ (\AA)&$\xi_{so}^{c}$(\AA)&$\xi_{so}^{ab}$ (\AA) &$\xi_{so}^{c}$ (\AA)&\\
\hline
LCCO& 115(12) & 59(2) & 195(4) & 22(1) & 98(8) & 22(1)\\
LSMO& $>$300 & `50 & $>$300 & `33 & &30(4)\\
LSCO& 23 & 8.3(6) &  &  & 79(3) & 10.7\\
LCNO& 30(10) & 2(1) &  &  & `120 & `13\\
\hline
\end{tabular}
\end{table}
\clearpage
\begin{table}[c]
\caption{Charge and magnetic correlation lengths of La$_{2-x}$Ca$_{x}$CoO$_{4}$. $\xi_{CO}$ and $\xi_{SO}$ are determined by the (0,1,7), (0,0.5,2.5) and (0.5,0,1) peaks. }
\label{t1}
\begin{tabular}{cccccccc}
\\
\hline
 &\multicolumn{2}{c}{$ $}&\multicolumn{2}{c}{$L=half-integer$}&\multicolumn{2}{c}{$L=integer$}\\
 &$\xi_{co}^{ab}$ (\AA)&$\xi_{co}^{c}$ (\AA)&$\xi_{so}^{ab}$ (\AA)&$\xi_{so}^{c}$(\AA)&$\xi_{so}^{ab}$ (\AA) &$\xi_{so}^{c}$ (\AA)&\\
\hline
0.3&  &  & 58(7) & 8(1) &  & \\
0.4& 50(9) & 17(1) & 34(2) & 15(1) & \\
0.5& 115(12) & 59(2) & 195(4) & 22(1) & 98(8) & 11(1)\\
0.6&    & 51(3) &  &  &  & 20(1)\\
0.5& 65(9) & 41(2) &  &  & 116(7) & 11(1)\\
0.5& 47(9) & 23(2) &  &  & 200(22) & 23(1)\\
\hline
\end{tabular}
\end{table}
\clearpage
\begin{table}[c]
\caption{Comparison between the calculated and observed magnetic scattering intensities in La$_{1.5}$Ca$_{0.5}$CoO$_{4}$. $\Phi$$_{s}$ is the angle between the spins and the (H,0,0) axis. $r$ indicates the type-I magnetic volume fraction.}
\label{t1}
\begin{tabular}{ccccc}
\\
\hline
 peak position&I$_{cal}$&I$_{obs}$&r$\cdot$I$_{cal}$&(0.5-r)$\cdot$I$_{cal}$\\ \hline
\hline
 (0.5,0,1)&268.3&58.5 & &49.6 \\
 (1.5,0,1)&188.0&35.6 & &34.8 \\
 (2.5,0,1)&104.1&15.6 & &19.3 \\
 (3.5,0,1)&48.6&6.4 & &9.0 \\
 (0,0.5,2.5)&341.5&93.6 &107.6 & \\
 (0,1.5,2.5)&196.3&47.5 & 61.8 & \\
 (0,2.5,2.5)&100.7&35.5 & 31.7 & \\
 (0.5,0,3)&323.3&62.1 & &59.8 \\
 (1.5,0,3)&192.3&32.3 & &35.6 \\
 (2.5,0,3)&97.8&18.3 & &18.1 \\
 (0.5,0,4)&273.3&44.9 & &50.6 \\
 (2.5,0,4)&173.6&38.2 & &32.1 \\
 (3.5,0,4)&72.0&6.2 & &13.3 \\
\hline
$\mu$=2.86(19)$\mu$$_{B}$, $\Phi$$_{s}$=48‹, $r$(type-I)=0.315
\end{tabular}
\end{table}
\clearpage
\begin{table}[c]
\caption{Comparison between the calculated and observed of the magnetic scattering intensities in La$_{1.85}$Ca$_{0.15}$CoO$_{4.17}$. $\Phi$$_{s}$ is the angle between the spins and the (H,0,0) axis. $r$ indicates the type-I magnetic volume fraction.}
\label{t1}
\begin{tabular}{cccc}
\\
\hline
 peak position&I$_{cal}$&I$_{obs}$&r$\cdot$I$_{cal}$\\ \hline
 (0,0.5,0.5)&354.5&177.8&177.3 \\
 (0,0.5,1.5)&533.5&236.9&266.8 \\
 (0,0.5,2.5)&541.1&272.0&270.5 \\
 (0,0.5,3.5)&474.3&251.9&237.1 \\
 (0,0.5,4.5)&386.8&200.9&193.4  \\
 (0,0.5,5.5)&300.4&149.9&150.2  \\
 (0,0.5,6.5)&224.6&117.3&112.3  \\
 (0,0.5,7.5)&165.0&72.1&82.5 \\
\hline
$\mu$=2.70(4)$\mu$$_{B}$, $\Phi$$_{s}$=38‹, $r$(type-I)=0.5
\end{tabular}
\end{table}
\clearpage
\begin{table}[c]
\caption{Comparison between the calculated and observed the magnetic scattering intensities. $\Phi$$_{s}$ is the angle between the spins and the (H,0,0) axis. $r$ indicates the type-I magnetic volume fraction.}
\label{t1}
\begin{tabular}{ccccc}
\\
\hline
 peak position&I$_{cal}$&I$_{obs}$&r$\cdot$I$_{cal}$&(0.5-r)$\cdot$I$_{cal}$\\ \hline
 (0.75,0.75,0.5)&67.7&22.8 &18.2 & \\
 (1.25,1.25,0.5)&46.0&10.3 &12.4 & \\
 (0.75,0.75,1)&66.1&13.1 & &13.9 \\
 (1.75,1.75,1)&42.9&9.3 & &4.5 \\
 (0.75,0.75,1.5)&63.0&15.5 &16.9 & \\
 (1.25,1.25,1.5)&42.9&9.9 & 11.5 & \\
 (0.75,0.75,2)&59.3&9.0 &  &10.2 \\
 (1.25,1.25,2)&26.2&4.9 & &7.0 \\
\hline
$\mu$=3.64(23)$\mu$$_{B}$, $\Phi$$_{s}$=90‹, $r$(type-I)=0.315
\end{tabular}
\end{table}

\end{document}